\documentclass[aps,floatfix,showpacs]{revtex4-1}
\usepackage{amsmath,amssymb,bm,graphicx,epsfig,psfrag,ulem,color,natbib}

\newcommand{\Reyn}{\mbox{\rm Re}_n}         

\newcommand{\bs}{\mathbf {s}}
\newcommand{\br}{\mathbf {r}}
\newcommand{\bv}{\mathbf {v}}

\newcommand{\etal}{{\it et al.}~}

\begin{document}

\title{Vortex density fluctuations in quantum turbulence}

\author{A.~W.~Baggaley}
\email{a.w.baggaley@ncl.ac.uk}
\author{C.~F.~Barenghi} 
\affiliation{School of Mathematics and Statistics, University of
Newcastle, Newcastle upon Tyne, NE1 7RU, UK}


\begin{abstract}
Turbulence in the low temperature phase of liquid helium is a
complex state in which a viscous normal fluid
interacts with an inviscid superfluid. In the former vorticity
consists of eddies of all sizes and strengths; in the latter vorticity 
is constrained to quantized vortex lines.
We compute the frequency spectrum of superfluid vortex 
density fluctuations and obtain the same
$f^{-5/3}$ scaling which 
has been recently observed. 
We show that the scaling can be interpreted in terms 
of the spectrum of reconnecting material lines. 
To perform this calculation we have developed a vortex tree algorithm 
which considerably speeds up the evaluation of Biot-Savart integrals.
\end{abstract}


\maketitle
Current theoretical and experimental work explores  
the relation between turbulence in an
ordinary (classical) fluid and turbulence in the quantum phases of 
$^4$He, $^3$He and atomic Bose-Einstein condensates. Quantum
turbulence\cite{Nemir}
 shares important  features with classical homogeneous isotropic
turbulence\cite{Vinen,Skrbek}: the most important is the
Kolmogorov $k^{-5/3}$ energy spectrum\cite{Tabeling}
where $k$ is the wavenumber. 

$^4$He consists of two components: an inviscid superfluid 
components (associated to the quantum ground state) and thermal excitations, 
which make up a viscous normal fluid component.  
What makes helium particularly interesting 
is that superfluid vorticity is concentrated in 
line singularities of fixed circulation $\kappa=h/m$, where $h$ is Planck's 
constant and $m$ is the mass of one helium atom (in $^3$He the relevant
boson is a Cooper pair).
Normal fluid vorticity is unconstrained, as in classical flows.
Turbulence in $^4$He is thus a complex doubly-turbulent regime, in which a  
viscous fluid interacts with discrete inviscid vortex lines.

The intensity of quantum turbulence is characterized by
the vortex line density $L$ (vortex length per unit volume). 
In a striking experiment,
Roche {\it et al.}\cite{Grenoble} measured 
the fluctuations of $L$ in turbulent $^4$He.
They observed that the frequency spectrum scales as 
$f^{-5/3}$, where $f$ is the frequency. The same scaling was observed
in turbulent $^3$He\cite{Lancaster}.
The rapid decrease of the spectrum is surprising because, if one
interprets $L$ as a measure of the rms superfuid vorticity
($\omega_s = \kappa L$), it seems to contradict the classical 
scaling of vorticity \cite{Farge} expected from the
Kolmogorov energy spectrum, which increases with $k$. The aim of this letter
is to shed light onto this problem.

Since the typical distance between superfluid vortices
$\ell \approx L^{-1/2}$ is orders of magnitude bigger than the vortex
core radius $a_0$, we  model vortices as space curves 
${\bf s}(\xi,t)$ where $\xi$ is arclength and $t$ is time. 
The curves are numerically discretised by a large, variable number of 
points $\bs_j$ ($j=1,\cdots N)$. 
The governing equation is \cite{Schwarz}

\begin{equation}
\frac{d\bs}{dt}=\bv_s+\alpha\bs' \times (\bv_n-\bv_s)
-\alpha'\bs' \times \left[ \bs' \times (\bv_n-\bv_s)\right],
\label{eq:Schwarz}
\end{equation}

\noindent
where $\bs'=d\bs/d\xi$ is the unit tangent vector at $\bs$,
$\alpha$, $\alpha'$ are temperature dependent friction coefficients
\cite{BDV-DB},
$\bv_n$ is the normal fluid's velocity, and the velocity $\bv_s$ which the
vortex lines induce on each other at the point $\bs$ is given by the 
Biot-Savart (BS) law:

\begin{equation}
\bv_s=-\frac{\kappa}{4 \pi} \oint_{\cal L} \frac{(\bs-\br) }
{\vert \bs - \br \vert^3}
\times {\bf d}\br.
\label{eq:BS}
\end{equation}
\noindent

\noindent
The line integral extends over the entire vortex configuration 
$\cal L$.  Vortex lines reconnect\cite{reconnections} when they 
become sufficiently close to each other,
provided that the total length (as a proxy
for energy) is reduced\cite{Leadbeater}. The discretization
technique is standard\cite{discretization}; the reconnection
technique and the de-singularization of the BS integral
are described elsewhere\cite{Baggaley}.

The difficulty of this vortex filament method is the computational cost 
of the BS law which scales as $N^2$ 
(the velocity at one point depends on an integral over
all other $N-1$ points); this prevents calculations of intense vortex tangles
(large $N$) for sufficiently long times to make realistic comparison with
experiments. The same difficulty arises in astrophysical
N-body simulations (the force of gravity
on one body depends on the other $N-1$ bodies); in this
context, the problem was solved by the development of tree 
algorithms \cite{Barnes} whose computational cost scale as 
$N\log(N)$ with small loss of accuracy \cite{Bertschinger,Barnes2}.

\begin{figure}
\begin{center}
\includegraphics[width=0.45\textwidth]{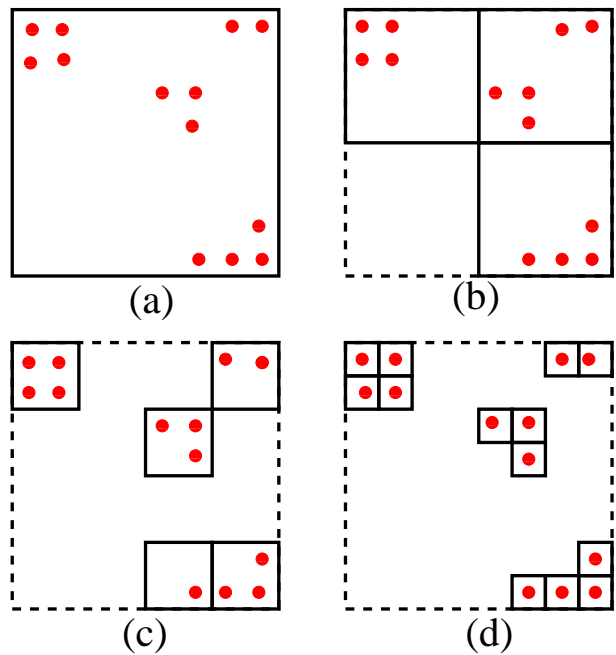}
\caption{
(Color online) Illustration of the tree construction 
in two dimensions (quad-tree).  
The points (red dots) are inclosed in the root cell (a), 
which is divided into four cells of half size (b), until (c,d) there
is only one point per cell.  
}
\label{fig1}
\end{center}
\end{figure}

To achieve our aim we have developed the following tree algorithm 
for vortex dynamics.  At each time step,  points are grouped in 
a hierarchy of cubes which is arranged in a three-dimensional 
oct-tree structure.
We construct the tree top down, first dividing the computational box 
(root) into eight cubes, and then continuing to divide each cube into eight 
`children', 
until either a cube is empty or only contains one point.
As we create the tree, we calculate the total 
vorticity contained within each cube and the 
`center of vorticity' of the cube  
from the points that it contains.
The time required for constructing the tree scales as $N\log(N)$, 
so it feasible to `redraw' the tree at each time step.
Fig.~\ref{fig1} illustrates this procedure in two dimensions.

To calculate the induced velocity $\bv_s$ at each point $\bs$ 
we must `walk' the tree, and decide if a cube is sufficiently far.
This is done using the concept of the opening angle $\theta$ 
\cite{Barnes} (as corrected by Barnes\cite{Barnes2,Dubinski} 
to avoid errors if the center of vorticity is near the edge of the cube).
Let $w$ be the width of the cube, $d$ the distance of the center of vorticity
from $\bs$ and $\delta$ the distance from the center of vorticity and 
the geometrical center of the cube. If $d>w/\theta+\delta$ we accept the cube,
and its contribution is used in computing the
velocity via Eq.~\ref{eq:BS}.  If not, then we open the cube
(assuming it contains more that one  point) 
and repeat the test on each of the child cubes that it contains. 
The tree-walk ends when the contributions of all cubes have been evaluated.

We tested the tree algorithm  
up to $N=2000$ points (practical limit of the BS law)
 using different values of 
$\theta$. We verified the $N\log(N)$ scaling for both 
the construction of the tree and the calculation of the 
total velocity. We found that the relative   
deviation of the velocity computed via the tree algorithm from the
exact BS velocity is at the most $0.25\%$
if $\theta=0.7$, which we take as the critical opening 
angle hereafter.
 
The computational box is a cube of size $D=0.075~\rm cm$ with periodic 
boundary conditions. When evaluating the BS integral, for each
point in the box we consider the other $3^3-1=26$ boxes around it;
this periodic wrapping is easily obtained
using the tree structure..

\begin{figure}
\begin{center}
   \includegraphics[width=0.45\textwidth]{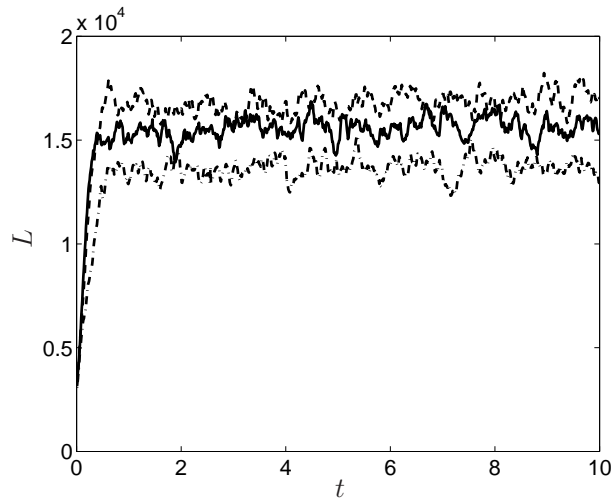}
   \caption{
   Vortex line density $L$ ($\rm cm^{-2}$) vs time $t$ ($\rm s$) 
   corresponding to
   $\Reyn=22.7$ (dot-dashed line), 
   $\Reyn=57.1$ (solid line), 
   and $\Reyn=112.9$ (dashed line). 
   }
\label{fig2}
\end{center}
\end{figure}
\begin{figure}
\begin{center}
   \includegraphics[width=0.45\textwidth]{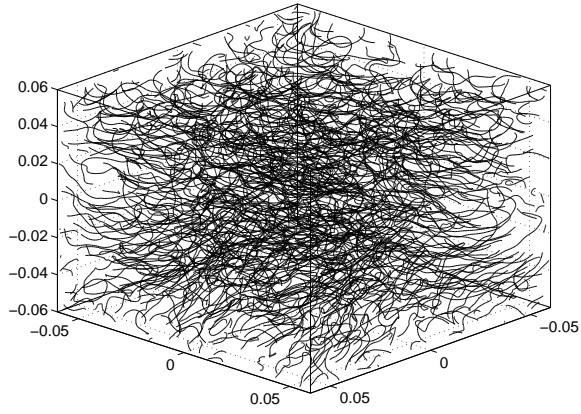}
   \caption{
    Saturated vortex tangle at $t=2.0~\rm s$ with $N=55,359$ and
    $L=91,733~\rm cm^{-2}$,
    corresponding to $\Reyn=507$.
    }
\label{fig3}
\end{center}
\end{figure}
\begin{figure}
\begin{center}
   \includegraphics[width=0.4\textwidth]{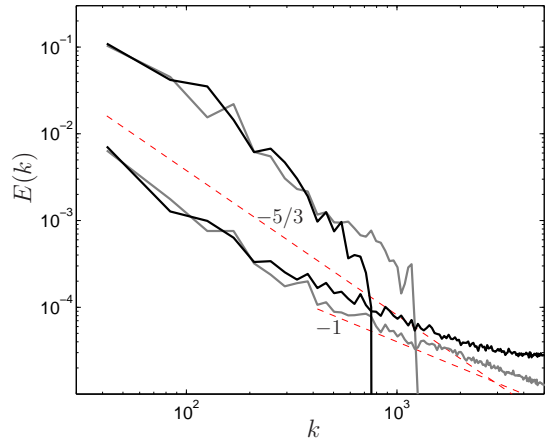}
   \caption{
   (Color online). Normal fluid's (upper two lines) and superfluid's
   (lower two lines) energy spectrum $E(k)$ vs $k$ ($\rm cm^{-1}$).
   Grey lines: $\Reyn=112.9$ ($L=8889~\rm cm^{-2}$,
   $k_{\ell}=2\pi/ \ell=592~\rm cm^{-1}$); 
   black lines: $Re_\textrm{n}=49.85$ ($L=7058~\rm cm^{-2}$,
   $k_{\ell}=527~\rm cm^{-1}$).
   Dashed lines:  $k^{-5/3}$ (top) and $k^{-1}$ (bottom) scalings.
   }
\label{fig4}
\end{center}
\end{figure}
\begin{figure}
\begin{center}
   \includegraphics[width=0.45\textwidth]{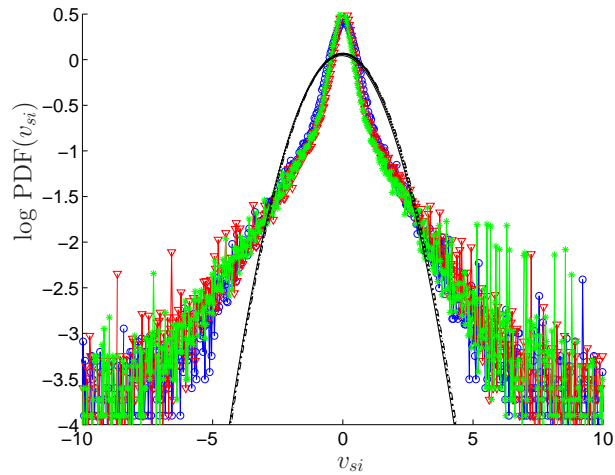}
   \caption{
   (Color online) PDF of superfluid velocity components
   ($\rm cm/s$) 
   $v_{sx}$ (blue circles), $v_{sy}$ (red triangles) 
   and $v_{sz}$ (green asterisks) sampled over the vortex-points
   for $\Reyn=112.9$.
   The overlapping black dotted, dash dotted and solid lines are respectively
   the Gaussian fits to the same data,
   $\text{gPDF}(v_{si}) =(1/(\sigma \sqrt{2\pi}))
   \exp{(-(v_{si}-\tilde{\mu})^2}/(2\sigma^{2}))$,
   ($i=1,2,3$) where $\sigma$ and $\tilde{\mu}$ are the
   standard deviation and the mean. 
   }
\label{fig5}
\end{center}
\end{figure}
\begin{figure}
\begin{center}
   \includegraphics[width=0.4\textwidth]{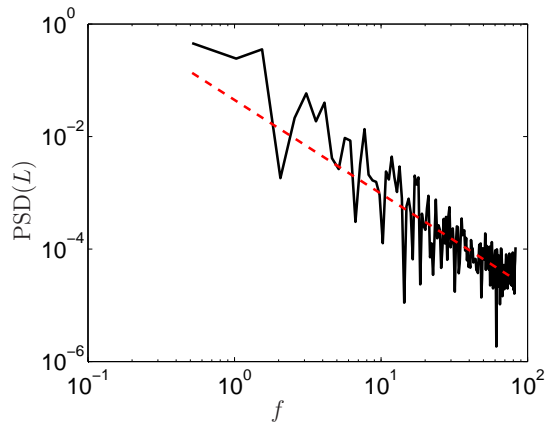}
   \caption{
   (Color online).
   Power spectral density of fluctuations of $L$ (arbitrary units)
   vs $f$  ($\rm s^{-1}$) 
   at $t=10~\rm s$
   corresponding
   to $\Reyn=507$ as in Fig.~\ref{fig3}. 
   The best fit to the data is $f^{-1.71}$. 
   The dashed line shows the $f^{-5/3}$ scaling. 
   }
\label{fig6}
\end{center}
\end{figure}
\begin{figure}
\begin{center}
   \includegraphics[width=0.4\textwidth]{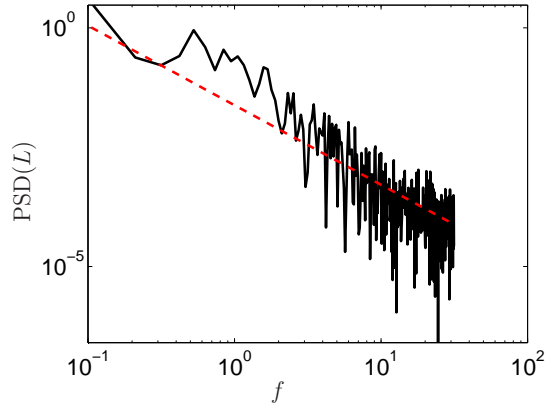}
   \caption{
   (Color online).
   Power spectral density of fluctuations of length of
   reconnecting material lines (arbitrary units)
   vs $f$  ($\rm s^{-1}$) at $t=10\rm s$,
   corresponding to $\Reyn=33.28$. 
   The best fit to the data is $f^{-1.74}$. 
   The dashed line shows the $f^{-5/3}$ scaling.
   }
\label{fig7}
\end{center}
\end{figure}

To model the turbulent normal fluid of the experiment\cite{Grenoble}
 we use a Kinematic Simulation (KS)
\cite{Osborne}, in which the normal fluid velocity at position $\bs$ and 
time $t$ is prescribed by the following sum of $M$ random Fourier modes:

\begin{equation}\label{eq:KS}
\bv_n({\bs},t)= \sum_{m=1}^M\left(\mathbf{A}_m \times \mathbf{k}_m
\cos\phi_m + {\bf B}_m \times \mathbf{k}_m \sin\phi_m \right),
\end{equation}

\noindent
with $\phi_m=\mathbf{k}_m \cdot \bs + \omega_m t$, where
$\mathbf{k}_m$ and $\omega_m=\sqrt{k_m^3 E(k_m)}$
are wavevectors and
frequencies.     
Via an appropriate choice of $\mathbf{A}_m$ and $\mathbf{B}_m$, 
the energy spectrum of $\bv_n$
reduces to the Kolmogorov form $E(k_m)\propto k_m^{-5/3}$ 
for $1\ll k\ll k_M$, with $k=1$ at the integral scale and $k_M$ 
at the cut-off scale. 
The effective Reynolds number 
 $\Reyn=(k_M/k_1)^{4/3}$
is defined by the
condition that the dissipation time equals the eddy turnover time at 
$k = k_M$.
Like some \cite{Wilkin} previous implementations of KS,
we have adapted Eq.~\ref{eq:KS} to periodic boundary conditions.
In summary, $\bv_n$ is solenoidal,
time-dependent, and satisfies the main
properties of homogeneous isotropic turbulence, from the 
energy spectrum to two-points statistics.

We use parameters which refer to $^4$He: circulation 
$\kappa=9.97 \times 10^{-4}~\rm cm^2/s$ and vortex core radius
$a_0=10^{-8}~\rm cm$. We choose $T=2.164~\rm K$ 
($\alpha=1.21$  and $\alpha'=-0.3883$), larger than in Ref.~\cite{Grenoble},
in order that
the back reaction of the vortex lines on the normal fluid 
is negligible and Eq.~\ref{eq:KS} is justified.
Our calculation has thus two independent parameters: $T$ and $\Reyn$.
The initial condition consists of 16 straight vortices at random positions 
and orientations.

Figure \ref{fig2} shows time series of the vortex line density at
three different values of $\Reyn$. In each case
the initial growth is followed by saturation to 
a statistical steady state in which $L$
fluctuates around a mean value. An example of a saturated
vortex tangle is shown in Fig.~\ref{fig3}. By harnessing the power of
the tree algorithm we have performed calculations with up to $N=400,000$
points.

We construct a $512^2$ mesh in the $xy-$plane 
at the center of the box. At each mesh point we calculate 
$\bv_s$ and $\bv_n$, using the tree approximation 
to the BS integral and Eq.~\ref{eq:KS} respectively. The corresponding
energy spectra for two values of $\Reyn$ are shown in Fig.~\ref{fig4}.
The $k^{-5/3}$ Kolmogorov spectrum of the normal fluid is clearly visible.
The superfluid follows the Kolmogorov scaling 
in the inertial range
$1\ll k\ll k_M$, in agreement with experiments\cite{Tabeling}. In the range
$k>k_M$ the normal fluid is essentially at rest, and the friction dampens 
Kelvin waves on quantized vortices (preventing a
cascade of energy to larger $k$ which would happen if $\alpha=\alpha'=0$,
as discussed in Ref.~\cite{Baggaley}); 
a $k^{-1}$ scaling, typical of
individual straight vortex lines, is visible in this range.

Despite the classical nature of the superfluid
energy spectrum, the statistics of superfluid velocity
components display power-law behavior. The probability
density functions (normalized histogram, or PDF for short)
scale as
${\rm PDF}(v_{s,i})\propto v_{si}^b$ ($i=1,2,3$)
with average exponent $b=-3.1$, see Fig.~\ref{fig5}. 
This scaling was observed in turbulent helium
experiments \cite{Paoletti}, and
was calculated in turbulent atomic
condensates\cite{White} and helium counterflow\cite{Adachi};
its cause is the singular nature of the superfluid vorticity\cite{White}.
The vortex line velocity $d\bs/dt$ obeys non-Gaussian scaling too.
The statistics of velocity components in
ordinary turbulence, on the contrary, are Gaussian\cite{classicalPDF}.

Finally we compute the frequency spectrum of the fluctuations 
of the vortex line density in the saturated state. 
Fig.~\ref{fig6} shows that the spectrum scales as $f^{-5/3}$, 
for large $f$, as observed in the experiments 
\cite{Grenoble,Lancaster}. The fact that the high frequency regime
where this scaling takes place is not exactly the same as in the
experiment is less important, and
arises from our choice of temperature, hence the value of saturated $L$.
We observe the same $f^{-5/3}$ scaling at all values of turbulent
intensities $\Reyn$.

What is the reason for this scaling ? Roche {\it et al.}
\cite{Roche-Barenghi,Salort}
argued that the more randomly oriented vortex lines 
(which particularly contribute to line length and second sound attenuation)
have some of the statistical properties of passive scalars. 
To test this idea we perform numerical simulations in which vortex
filaments are replaced by passive material lines which evolve
according to $d\bs/dt=\bv_n$
(we do not switch off the reconnection algorithm, otherwise the vortex 
length would grow indefinitevely).
We find that the length saturates at values larger
than the vortex line density 
and that the spectrum of the fluctuations scales again
as $f^{-5/3}$, as in Fig.~\ref{fig6}. 
Being $L$ positive definite\cite{Ishihara},
there is no conflict with the vorticity spectrum.

In conclusion, our calculations reproduce the main observed features
of quantum turbulence: (i) the classical
$k^{-5/3}$ scaling of the energy spectrum observed at large scales 
by Tabeling\cite{Tabeling} 
(thought to be
associated with large-scale, energy-containing 
polarization of vortex lines\cite{Vinen});
(ii) the observation of non classical (non Gaussian) velocity
statistics\cite{Paoletti} 
(macroscopic manifestation of singular vorticity\cite{White}); and
(iii) the
$f^{-5/3}$ spectrum of the vortex line density fluctuations observed
at large frequency\cite{Grenoble,Lancaster}. The natural question
is whether they are independent.
Our results also support Roche's 
interpretation\cite{Roche-Barenghi} that vortex density fluctuations
arise from random vortex lines which behave
as reconnecting material lines (while most of the tangle's energy is in
the large scale motion). 

The vortex tree algorithm could be further speeded up
by parallelization\cite{Springel}. Its power
should allow us to tackle problems which require large $N$, for example
the detection of anomalous scaling\cite{Frisch}
and, in the $T=0$ limit, the transition from the Kolmogorov 
energy cascade at small $k$ to the Kelvin waves cascade at big $k$
\cite{Lvov-Kozik}.



\begin{thebibliography}{99}
\bibitem{discretization}
During the evolution new points are introduced
(using quadratic interpolation) if the distance between two 
neighboring points along a vortex becomes larger than 
$\Delta \xi_{min}=0.001~\rm cm$.  
Similarly points are removed, keeping their
mutual distance comprised between $\Delta \xi_{min}$ and
$\Delta \xi_{min}/2$.
Time evolution is based on a $3^{rd}$ order Adams-Bashforth scheme
(the time step $\Delta t$ depends on $\Delta \xi_{min}$ to resolve
the shortest Kelvin waves).
\end{thebibliography}
\end{document}